\documentclass[10pt]{article}

\usepackage{amsmath}
\usepackage{array}
\usepackage{appendix}
\usepackage{graphicx}
\usepackage{amsfonts}
\usepackage{amssymb}

\def\be{\begin{equation}}
\def\ee{\end{equation}}
\def\beq{\begin{equation}}
\def\eeq{\end{equation}}
\def\bea{\begin{eqnarray}}
\def\eea{\end{eqnarray}}

\def\ni{\noindent}

\def\hat{\widehat}

\def\!{\hspace{-1.6667em}}

\def\mC{\mbox{C}}

\def\mS{\mbox{S}}

\def\mV{\mbox{V}}

\def\mh{\mbox{h}}

\def\bupSigma{\mbox{\boldmath$\Sigma$}}                 

\def\sd{\mbox{\scriptsize d}}

\def\sf{\mbox{\scriptsize f}}

\def\si{\mbox{\scriptsize i}}

\def\sn{\mbox{\scriptsize n}}

\def\st{\mbox{\scriptsize t}}

\def\pa{\partial}
\def\d{\textrm{d}}

\def\5Star{\mbox{\Large$\star$}}              
\def\cr{\mbox{\scriptsize{\bf $\mbox{ } \times \mbox{ }$}}}

\def\sumi2{\sum\mbox{}_{\mbox{}_{\mbox{\scriptsize $i$=1}}}^2}
\def\sumi3{\sum\mbox{}_{\mbox{}_{\mbox{\scriptsize $i$=1}}}^3}

\def\sumj3{\sum\mbox{}_{\mbox{}_{\mbox{\scriptsize $j$=1}}}^3}
\def\sumk3{\sum\mbox{}_{\mbox{}_{\mbox{\scriptsize $k$=1}}}^3}

\begin{document}

\begin{center}

{\bf \Large `SHAPE DYNAMICS': FOUNDATIONS REASSESSED}

\vspace{.15in}

{\large \bf Edward Anderson} 

\vspace{.15in}

\large {\em DAMTP, Centre for Mathematical Sciences, Wilberforce Road, Cambridge CB3 OWA.  } \normalsize

\end{center}

\begin{abstract}

`Shape dynamics' is meant here in the sense of a type of conformogeometrical reformulation of GR, 
some of which have of late been considered as generalizations of or alternatives to GR.
This note concerns in particular cases based on the notion of volume-preserving conformal transformations (VPCTs), 
in the sense of preserving a solitary global volume of the universe degree of freedom. 
The extent to which various ways of modelling VPCTs make use of group theory at all, in a congruous manner, 
and with minimal departure from standard Lie group theory, is considered. 
This points to changing conception of VPCTs from the current finite integral implementation to an infinitesimal differential implementation (or to avoiding using them at all).  
Some useful observations from flat-space conformal groups (well-known from CFT) concerning the existence or otherwise of VPCT groups are also provided. 
 
\end{abstract}

\section{Introduction}\label{Introduction}

A number of relational models -- in the Leibnizian \cite{L} and Machian \cite{M} senses of the word -- are now available 
\cite{RWR-AM13, BO, ABFO, B03-FileR-ASphe, Mercati14, AConfig, AMech}. 
These trace their roots back to \cite{BB82}. 
Almost all of these relational models implement infinitesimal symmetry transformations as indeed is customary more widely in both Physics and Mathematics (theory of Lie groups \cite{Lie}).
The conformal `CS + V' \cite{ABFKO} and `shape dynamics' \cite{SD, Mercati14} cases are exceptions in this sense, 
due to which the current note places their mathematical and physical foundations under scrutiny.  
Some further motivation for these relational formulations and theories are approaches viewing these that as reformulations of, or alternatives to, GR. 
They are also of further value as regards study of aspects of Background Independence and of ensuing Problem of Time facets \cite{PoT}.  
Additionally, conformal formulations of GR are also of major importance in the GR initial-value problem (IVP) \cite{York72, IVP}.

\section{Four implementations of volume-preserving conformal transformations}

Begin by considering the general conformal transformation (CT) in the presentation 
\beq
h_{ab} \longrightarrow \phi^4 h_{ab}
\label{CT} 
\eeq
which is convenient in 3-$d$: the spatial dimension. 
Then also 
\beq
\sqrt{h} \longrightarrow \phi^6\sqrt{h} \mbox{ } .
\eeq
The CTs form the group well-known group \cite{FM96} $Conf(\bupSigma)$; here $\bupSigma$ is the 3-$d$ space's topology, 
taken in this article to be a fixed instance of compact without boundary manifold.

\cite{BO, ABFO} just made use of no more than (\ref{CT}). 
However, the full CTs correspond to maximal slices\footnote{$p = p^{ij}h_{ij}$ for $h_{ij}$ the spatial 3-metric 
and $p^{ij}$ the GR momentum conjugate to that.}
\beq
p = 0 \mbox{ } 
\label{max}
\eeq 
[indeed actions built to relationally incorporate (\ref{CT}) encode (\ref{max})].
It is then well-known that maximal slices cannot then be continued so as to form a foliation of such in GR with closed $\bupSigma$.

One way of avoiding this conclusion is to divide the action by the homogenizing power of the volume of the universe, $\mV(\bupSigma)$.
However, this manoeuvre -- similar to one used in work on the Yamabe Conjecture \cite{Yamabe-Schoen} that is familiar from the GR IVP literature -- 
in this case leads to an alternative theory in place of GR, with action at a distance and no apparent means of having a viable cosmology \cite{ABFO}.

Thus a second way of avoiding the above conclusion was considered, 
now paralleling York's generalization \cite{York72} of Lichnerowicz's work \cite{Lich44} on the GR IVP: from maximal slices to CMC slices.  
This stands for `constant mean curvature': 
\beq
const = p/\sqrt{h} \propto K 
\label{CMC}
\eeq
for $K = K_{ij}h^{ij}$ and $K_{ij}$ the extrinsic curvature of space within spacetime.
The approach to conformal relationalism along these lines then proceeded by considering (global) volume-preseving conformal transformations (VPCTs).

\mbox{ }

\ni The implementation\footnote{`fin-int' stands for finite integral implementation; 
$\hat{\phi}$ is the notation used in the literature so far.  
The `inf' and `diff' subscripts used below stand for infinitesimal and differential respectively.
$\langle O \rangle$ is the volume-average $\int_{\Sigma} O\sqrt{h} \, \d^3x/\int_{\Sigma} \sqrt{h} \, \d^3x$ of some object $O$.    
}
of these used in \cite{ABFKO} -- and usually used since in the `shape dynamics' literature -- is
\be
\phi^{\sf\si\sn-\si\sn\st} \mbox{ } (  \mbox{ } = \mbox{ } \hat{\phi} \mbox{ }  ) \mbox{ } = \mbox{ } \frac{\phi}{\langle\phi^6\rangle^{1/6}} \mbox{ } . 
\label{fin-int}
\ee  
This assigns a VPCT to each CT, as is clear upon integration of the VPCT transformed volume element $\hat{\phi}^6\sqrt{h}$ over space $\bupSigma$ 
returning the same global volume as that from integrating the conformally untransformed volume element $\sqrt{h}$.

\mbox{ } 

\ni 1) Note however that this is indeed a finite rather than infinitesimal transformation.
In this way, it is unlike the other transformations used in the relational models that trace their roots back to \cite{BB82}; 
these transformations which are infinitesimal translations, rotations, scalings, spatial diffeomorphisms etc, entering by the so-called `Best Matching' technique \cite{BB82, Mercati14}.  

\ni 2) Note furthermore that the $\hat{\phi}$ do not even close as a group: $\hat{\phi_1}\hat{\phi_2}$ is not itself of the form (\ref{fin-int}).  
This is due to the products and integrals not being in the right order: 
$$
\phi^{\sf\si\sn-\si\sn\st}_1\phi^{\sf\si\sn-\si\sn\st}_2 = \frac{\phi_1\phi_2}{(\langle\phi_1^6\rangle \langle\phi_2^6\rangle)^{1/6}} \neq 
                                                           \frac{\phi_1\phi_2}{\langle(\phi_1\phi_2)^6\rangle^{1/6}}                        \mbox{ } .
$$ 
Thus whereas (\ref{fin-int}) implements {\sl individual} VPCTs, it does {\sl not} implement VPCTs {\sl as a group}.

\mbox{ }

\ni Moreover, need VPCTs even exist as a group?   
A brief digression to flat spacetime, flat space and maximally-symmetric spaces more generally -- 
a topic well-known from the study of CFT\footnote{See \cite{Kos15} for other inter-plays between shape dynamics and CFT.} -- is useful here.  
On the one hand, in flat spacetime or space, the special conformal transformations $K_i$ do not close without the overall dilation $D$ \cite{AMP-West}. 
This is due to their Lie bracket with the translations $P_i$ producing both $D$ and $d$-dimensional rotations $M_{ij}$ as integrabilities.  
This suggests that caution is needed as regards whether well-defined global scale to VPCT splits exist. 
That is not to be presupposed, but rather to be checked within the framework of Group Theory.   
{\sl Splitting CTs into global scalings and VPCTs may sound innocuous, but the VPCTs are capable of not being able to stand alone on group-theoretic grounds.}  
On the other hand, for the $p$-sphere of fixed radius, scale is meaningless but conformal transformations indeed still exist \cite{G03}. 
Thus in this setting VPCTs both exist and are all of the setting's CTs. 
I will next consider three ways around 1) and 2).  

\mbox{ } 

\ni A) Use an infinitesimal version of the VPCT implementation instead. 
Expanding out, this takes the form 
\be
\phi^{\si\sn\st-\si\sn\sf} = 1 + \xi - \langle{\xi}\rangle =: 1 + \overline{\xi} \mbox{ } , 
\label{inf-int}
\ee
i.e. correction by a small contrast-type object.  
These do indeed close as a group: $1 + \overline{\xi_1} + \overline{\xi_2} = 1 + \overline{\xi_1 + \xi_2}$.  
Whenever we succeed in forming a group of VPCTs, we denote it $VPConf(\bupSigma)$.

\mbox{ } 

\ni 3) However, even this is not within the scope of Lie groups, due to its integro-differential character. 
Indeed, Lie group generators are {\sl differential} in nature, 
e.g. $P_i = - \pa/\pa x^i$ for the translations or $M_{ij} = x_i\pa/\pa x^j - x_j\pa/\pa x^i$ for the $d$-dimensional rotations).
E.g. the usual `Euclidean' relational particle mechanics' corresponding Best Matching correction to the change ($\d$) in particle configuration $\underline{q}^I$ is $\d \underline{q}^I 
- \d \underline{a} - \d \underline{b} \cr \underline{q}^I$ (for particle positions $\underline{q}^I$ and translation and rotation auxiliaries $\underline{a}$ and $\underline{b}$.
E.g. also GR as relational geometrodynamics' is $\d h_{ab} - 2D_{(a}\d F_{b)}$ 
(for frame variable $F_{b}$ which is closely related to the GR shift vector, and $D_a$ the spatial covariant derivative). 
Contrast these examples with (\ref{fin-int})'s 
\beq
\phi^{\sf\si\sn-\si\sn\st \, 4}
\left(
\d h_{ab} + h_{ab}
\left(
4\left(
\frac{\d \phi}{\phi} - \frac{\langle \phi^5 \d\phi \rangle}{\langle \phi^6\rangle}
\right)
+ \frac{1}{3}
\left(
\langle  h^{ab}\d h_{ab} \rangle - \frac{\langle  \mh^{ab}\d h_{ab} \phi^6 \rangle }{\langle  \phi^6 \rangle}
\right)
\right)
\right)
\label{phi-corr}
\eeq
which contains both integral signs and further involvement of the change of configuration itself  -- $\d \mh_{ab}$ -- within the correction terms. 
Passing to the infinitesimal form does not alleviate these features: 
\beq
(1 + \bar{\xi})^4
\left(
\d h_{ab} + h_{ab} \, \frac{4 \d\xi + \langle 2 \xi \mh^{ab}\d h_{ab}  + \d \xi \rangle  - 2 \langle h^{ab} \d h_{ab}  \rangle \langle \xi \rangle}{1 + \bar{\xi}} 
\right)
\eeq

Furthermore, both the finite and the infinitesimal integral implementation give rise to the 
\beq
\overline{p/\sqrt{h}} = p/\sqrt{h} - \langle p/\sqrt{h} \rangle = 0 
\label{barp}
\eeq
form of CMC condition, corresponding to a combination of generators 
\beq
\overline{h_{ab}\frac{\delta}{\delta h_{ab}}} = h_{ab}\frac{\delta}{\delta h_{ab}} - \left\langle h_{ab}\frac{\delta}{\delta h_{ab}}\right\rangle
\eeq
which, by containing integral signs, is outside of the remit of Lie algebra generators.

\mbox{ } 

\ni B) Another way around 1) and 2) which comes {\sl closer} to resolving 3) as well is to use the infinitesimal differential Laplacian implementation.

\mbox{ } 

\ni Let us firstly recollect the finite version of this \cite{LanThan}:
\be
h_{ab} \longrightarrow \phi^{\sf\si\sn-\sd\si\sf\sf\,4}h_{ab} = (1 + D^2\zeta)^{2/3} h_{ab}
\label{fin-diff}
\ee
These indeed implement VPCTs due to 
$$
\int_{\Sigma}\phi^{\sf\si\sn-\sd\si\sf\sf \, 6}\sqrt{h} \, \d^3x = \int_{\Sigma}(1 + D^2\zeta) \sqrt{h} \, \d^3x = \int_{\Sigma}\sqrt{h} \, \d^3x
$$
using that $\bupSigma$ has no boundary.

This then encodes 
\beq
D^2 (p/\sqrt{h}) = 0 \mbox{ } , \mbox{ } \mbox{ among the solutions of which are the CMC slices $p/\sqrt{h} = const$ } \mbox{ } . 
\label{D^2p}
\eeq
Indeed, this is similar to the manner in which the conformal relationalism option arises 
by application of the Dirac algorithm \cite{Dirac} to geometrodynamical ans\"{a}tze more general than GR's own geometrodynamics \cite{RWR-AM13}.

Moreover, (\ref{fin-diff}) once again do not close as a group.\footnote{In fact, this was noted at the time by both the Author and \'{o} Murchadha for the differential implemenation.
The current note, however, points out that this problem holds in the integral implementation used since as well.}
%
Additionally, the corresponding generators are of the form 
\be
D^2
\left( 
h_{ab}\frac{\delta}{\delta h_{ab}}
\right) \mbox{ } .
\eeq

\ni Finally, expanding out, the corresponding infinitesimal version is then \cite{LanThan} 
\be
h_{ab} \longrightarrow \phi^{\si\sn\sf-\sd\si\sf\sf \,4}h_{ab} = \left(1 + \mbox{$\frac{2}{3}$}D^2\chi\right) h_{ab} \mbox{ } .
\label{inf-diff}
\ee
This now indeed also manages to close as a group: $1 + \mbox{$\frac{2}{3}$}D^2\xi_1 + \mbox{$\frac{2}{3}$}D^2\chi_2 = 1 + \mbox{$\frac{2}{3}$}D^2(\chi_1 + \chi_2)$,
while still encoding (\ref{D^2p}).  

\mbox{ }

\ni C) One can also consider group actions other than just multiplicative ones, as done in \cite{GT}.
Note moreover that this was not considered in setting up the first published works in this subject \cite{ABFO, ABFKO}, 
and that A) and B) above remain as alternatives in developing this subject.

\section{Conclusion and commentary}

\ni In summary, closing as a group, and being infinitesimal transformations like all the other continuous transformations involved in these theories, 
points to use of (\ref{inf-int}) or (\ref{inf-diff}). 
{\sl Neither} of these is Lie.  
(\ref{inf-diff}) has the advantage of only being out from this by the presence of higher derivatives rather than of integrals. 
This may well be closer to the spirit of standard Lie theory.  
Also GR's Dirac algebroid of constraints \cite{Dirac, BojoBook} already involves additional differential operators, 
so choosing the differential option involves staying within a complication which GR itself already manifests.

\mbox{ } 

\ni Some further significance of the current note is as follows. 

\ni I) Whichever of passing to an infinitesimal presentation and using a differential rather than integral implmentation changes the 
entirety of the detailed calculations of the ensuing shape dynamics (i.e. the ones explicitly involving $\phi$'s). 
On the other hand, the `core $\phi = 1$ equations' of the schemes, such as the Hamiltonian constraint, momentum constraint and CMC slicing condition, remain coincindent.  
This means that all of the schemes in question make contact with CMC-sliced GR paralleling formulations used in the GR IVP \cite{York72, IVP}. 

\ni II) Whenever any non group forming conception of VPCTs is in use, 
it then furthermore does not make sense to consider spaces obtained by `quotienting out that implementation of VPCT'.
Thus viewing\footnote{This is for $\mbox{Riem}(\bupSigma)$ the space of spatial 3-metrics, 
$Diff(\bupSigma)$ the corresponding spatial diffeomorphisms and $\rtimes$ denoting semi-direct product.}
the `conformal superspace + volume' space $(\mC\mS + \mV)(\bupSigma)$ as 
$\mbox{Riem}(\bupSigma)/VPConf(\bupSigma) \rtimes Diff(\bupSigma)$ can be more contentious than viewing it as 
$(\mbox{Riem}(\bupSigma)/Conf(\bupSigma) \rtimes Diff(\bupSigma)) + \mV(\bupSigma)$, i.e. adjunction of $\mV(\bupSigma)$ to a quotient by a bona fide group.  
Note furthermore that the latter is indeed how York himself originally presented $(\mC\mS + \mV)(\bupSigma)$.  

\ni III) Also note that none of the implementations of VPCTs used in this paper reduce to those of the $p$-sphere model, for which standard Lie mathematics is entirely sufficient.   
This gives a further `non-coincidence with simple and well-known limiting cases' type concern with existing implementations of VPCTs in general conformogeometrodynamical 
theories and formulations.
This would be even more of a concern if one were to step outside this note's compact without boundary remit, 
since in flat space and flat spacetime the conformal transformations excluding the constant dilation do not algebraically close.  

\ni IV) If `symmetry trading' is to be applied to `shape dynamics', 
one had better check that the {\sl notion of symmetry} involved in such trading corresponds to that manifested by the actual theory in question's generators.
These lying well outside of the usual Lie theory, it is far from clear whether such a generalized sense of symmetry lies within the scope of current treatises on `symmetry trading'.
Certainly the most commonly encountered examples of symmetry trading in Gauge Theory \cite{HT} do not by themselves suffice as a guarantee 
of being able to `trade' the far more general and far less explored notion of symmetry pointed to in this note. 
Having by now searched the literature quite extensively, 
I have not found evidence for the mathematical justification of `symmetry trading' having been generalized from conventional Lie algebra based gauge theories to more general structures 
such as the algebroids encountered in canonical study of gravitational theories such as GR.  

\ni V) Note that in addition to the previous Sec's `two ways around' the initial obstruction, Barbour and \'{o} Murchadha more recently provided a third way \cite{BO10}.
This is based on reinterpreting York's use of $(\mC\mS + \mV)(\bupSigma)$ as containing a global scaling redundancy, 
by which CMC mathematics is argued to be tied, rather, to $\mC\mS(\bupSigma)$.
Since this approach does not introduce VPCTs, it avoids the current note's problems.  
Consequently, we suggest more work be done in this specific direction.  

\mbox{ } 

\ni{\bf Acknowledgements} Thanks to those close to me. 
Thanks also to those who hosted me and paid for the visits: Jeremy Butterfield, John Barrow and the Foundational Questions Institute.
Thanks also to Julian Barbour, Niall \'{o} Murchadha, Brendan Foster, Flavio Mercati, Tim Koslowski and Chris Isham for a number of discussions over the years.


\end{document}